# Direct Time-Domain Observation of $l$-Doubling via Centrifugal-Distortion Pre-compensation

Inbar Sternbach[1,2], Kfir Rutman Moshe[1,2], Amit Beer[1,2], Soumitra Hazra[1,2], and Sharly Fleischer*[1,2]

[1]*Raymond and Beverly Sackler Faculty of Exact Sciences, School of Chemistry, Tel Aviv University 6997801, Israel.*
[2]*Tel-Aviv University center for Light-Matter-Interaction, Tel Aviv 6997801, Israel*
**Email: sharlyf@tauex.tau.ac.il*

**Abstract**

We demonstrate direct time-domain observation of $l$-doubling contributions in molecular rotational dynamics using shaped femtosecond laser pulses. By imposing a tailored spectral phase on the excitation pulse, we pre-compensate centrifugal distortion, which otherwise leads to temporally broadened, multi-cycle revival structures that obscure fine rotational features. A cubic spectral phase [1] compresses selected revivals into near single-cycle events, in agreement with an analytic expression derived from molecular rotational constants, enabling predictive pulse design beyond numerical optimization. The resulting distortion-free revivals reveal temporally separated $l$-doubling contributions that remain unresolved in conventional impulsive alignment experiments. The method proves robust against experimental imperfections, including spatial light modulator discretization. While selective control of individual $l$-doubling components becomes feasible, here we focus on their direct observation in the time domain.

---

Laser-induced rotational wave packets in gas-phase molecules exhibit periodic rephasing known as quantum rotational revivals [2–5]. These revivals underpin a wide range of applications, including rotational coherence spectroscopy and molecular-frame measurements such as high-harmonic generation and ultrafast diffraction [2–9]. In an ideal rigid rotor, the pure harmonic structure of rotational energy levels leads to perfectly periodic revivals. In real molecules however, centrifugal distortion introduces a $J^4$ correction to the rotational energies, resulting in dephasing that manifests as temporally broadened, multi-cycle revival structures [10–12]. Time-domain resolution of closely spaced rotational contributions, e.g. molecular isotopologues [13], relies on their gradual desynchronization. Centrifugal distortion (CDN) allows this desynchronization but simultaneously broadens the revival signals, maintaining their overlap. Rotational Echo schemes reverse the CDN dephasing but refocus all contributions to a common time [14–17], precluding their selective resolution

Here we demonstrate direct time-domain resolution of $l$-doubling contributions in molecular rotational dynamics by pre-compensating centrifugal distortion at the excitation stage. $l$-doubling, arising from rovibrational coupling in degenerate bending modes, is well established in frequency-domain spectroscopy [18–21], yet has only rarely been addressed in alignment experiments, with only a few reports in the literature [22,23], none of which resolved its alignment signature in the time domain. The limited visibility of $l$-doubling arises from the interplay between CDN-induced broadening and the need for long-time desynchronization of closely spaced rotational contributions. By tailoring the excitation pulse to cancel centrifugal dephasing at a chosen delay, we generate temporally compressed revivals that preserve this desynchronization, enabling separation of otherwise overlapping contributions. Applying this approach to $CO_2$, we resolve the distinct revival signals associated with $l$-doubling directly in the time domain.

In linear triatomic molecules such as $CO_2$, $l$-doubling results in the splitting of rovibrational levels into closely spaced components. In the time domain, this manifests as multiple rotational wavepacket contributions with slightly different rotational constants, corresponding to the ground vibrational state and the excited bending mode. The resulting alignment signal is therefore a coherent superposition of revivals with nearly identical (<1% difference) periodicities. While these components gradually desynchronize over time due to their small frequency mismatch, CDN simultaneously broadens each revival, maintaining significant temporal overlap and obscuring their individual contributions.

We first briefly describe the experimental approach. We then present an analytical framework for designing the spectral phase required to compensate CDN. This approach is validated experimentally in $CH_3I$, a system with strong centrifugal distortion but without $l$-doubling, and subsequently applied to $CO_2$ to resolve the distinct rotational contributions associated with $l$-doubling.

## Experimental setup

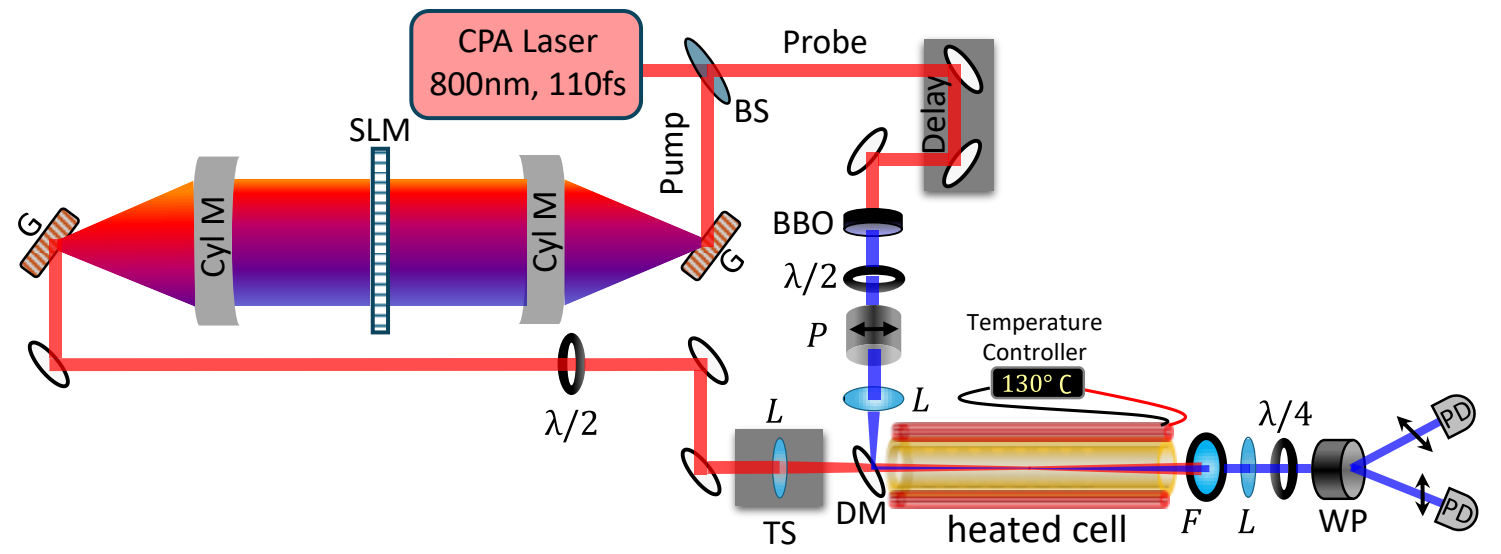

Figure 1: Transient birefringence setup. $P$ – polarizer, L- lens, $\lambda/2$-half wave plate, $\lambda/4$ – quarter wave plate, BBO crystal, WP- Wollaston prism, TS – Translation stage, F – filter (BG40), PD- Potodetector. The pump beam is routed through a pulse shaper, 4-f setup where G-grating, Cyl M are cylindrical reflective lenses and SLM- Spatial Light Modulator (dual mask, 640 pixels, Jenoptik).

We monitor the anisotropic angular distribution of rotationally excited gas phase molecules in the gas phase using time-resolved optical birefringence measurements [24,25] in collinear configuration [12,14,15,23,26,27]. An ultrashort laser beam (~110fs, 800nm) is split into two arms: a pump beam for rotational excitation (90%) at the fundamental frequency and weak probe (10%) beam that is frequency doubled in a BBO crystal to form a weak 400nm readout pulse (probe). The pump beam is routed through a 4-f pulse shaper system with two gratings and a spatial light modulator (Jenoptik) at the Fourier plane [28] for phase shaping. The probe polarization is set at 45$^{\circ}$ relative to the shaped pump and the polarization state of the transmitted probe is analyzed after the sample using a λ/4 plate, a Wollaston prism and balanced photodetection to measure the induced birefringence.

## Molecular alignment via stimulated Raman excitation

Rotational excitation by nonresonant ultrashort pulses arises from the interaction between the field intensity and the molecular polarizability anisotropy. The Hamiltonian is given by $H = \frac{\hat{L}^2}{2I} - \frac{1}{4}\Delta\alpha|E(t)|^2\cos^2\theta$, where $\hat{L}$ is the angular momentum operator, $I$- molecular moment of intertia, $\Delta\alpha$ is the polarizability anisotropy, $|E(t)|^2$ is the pulse intensity envelope and $\theta$ is the angle between the molecular axis and the laser polarization. Linear molecules are modeled as rigid rotors, with eigenstates $|J, m\rangle$ for the spherical harmonics functions and eigenenergies $E_J = BJ(J+1) - DJ^2(J+1)^2$, where the second term accounts for CDN. The interaction with the pulse creates a coherent rotational wavepacket, $|\psi\rangle(t) = \sum_{J,m} c_{J,m} \cdot \exp(-E_J t/\hbar) \cdot |J,m\rangle$

For linearly polarized excitation, the selection rules are $\Delta J = \pm 2, \Delta m = 0$. Since the coherent phase factor depends solely on $J$ we can restrict our analysis to the $J$ quantum number for brevity.

The time-dependent alignment is given by

$$\langle \cos^2\theta \rangle(t) = \sum_J c_J^* c_{J+2} \langle J|\cos^2\theta|J+2\rangle e^{-i\phi_{rigid}(J,t)} e^{i\phi_{CDN}(J,t)} \quad (1)$$

where $\phi_{rigid}(J,t) = 4\pi Bc(2J+3)t$ is the rigid rotor phase that gives rise to periodic revivals with period $T_{rev} = (2Bc)^{-1}$. At integer revival times, $t = nT_{rev}$, the rigid-rotor contribution rephases across all $J's$, and the alignment reduces to

$$\langle \cos^2\theta \rangle(n) = \sum_J c_J^* c_{J+2} \cdot \langle J|\cos^2\theta|J+2\rangle\, e^{i\phi_{CDN}(J,n)} \quad (2)$$

where the residual phase $\phi_{CDN}(J,n) = 4\pi n \frac{D}{B}(2J^3 + 9J^2 + 15J + 9)$ is dominated by CDN.

*Our goal is to shape the excitation pulse such that the phase imprinted on the Raman coherences,* $\arg(c_J^* c_{J+2})$*, cancels* $\phi_{CDN}$ *at a chosen revival, thereby suppressing distortion and restoring temporally compressed revivals.*

Shaping the phase of rotational Raman coherences

The rotational coherence amplitudes $(c_J^* c_{J+2})$ are induced via stimulated Raman transitions, driven by pairs of frequency components within the bandwidth of the excitation pulse. In order to calculate $c_J^* c_{J+2}$ with frequencies $\Omega = (E_{J+2} - E_J)/\hbar$, we must integrate over all of the frequency pairs that contribute to a Raman frequency $\Omega$:

$$R(\Omega) \propto \int \mathcal{E}(\omega) \cdot \mathcal{E}^*(\omega - \Omega) d\omega \quad (3)$$

where $\mathcal{E}(\omega) = |\mathcal{E}(\omega)| \cdot e^{i\Phi_{shape}(\omega)}$ is the spectral amplitude of the excitation pulse, $R(\Omega)$ is the complex coherent amplitude $(c_J^* c_{J+2})$ , and the Raman phase is: $\phi_{Raman}(\Omega) = \arg(R(\Omega))$. We consider an excitation pulse with a Gaussian spectral envelope:

$$|\mathcal{E}(\omega)| = \exp[-(\omega - \omega_0)^2/(2\sigma^2)]$$

Note that the CDN phase $(\phi_{CDN})$ is a polynomial of 3rd order in $J$. Discarding the linear term in $J$ which merely results in a shift in time and the ($J$-independent) global phase term, we restrict our pulse shaping efforts to cubic and quadratic terms:

$$\Phi_{shape}(\omega) = \tilde{a}(\omega - \omega_0)^2 + \tilde{b}(\omega - \omega_0)^3 \quad (4)$$

where $\tilde{a}$ , $\tilde{b}$ are the coefficients of the quadratic and cubic terms of the shaped near-IR pulse. Plugging $\mathcal{E}(\omega)$ into the integral (eq.3), the resulting Raman coherences are expressed as:

$$R(\Omega) \propto \int \exp\left[-x^2\left(\frac{1}{\sigma^2} - i3\tilde{b}\Omega\right) + x\left(\frac{\Omega}{\sigma^2} + i\left(2\tilde{a}\Omega - 3\tilde{b}\Omega^2\right)\right) + i\left(\tilde{b}\Omega^3 - \tilde{a}\Omega^2\right) - \frac{\Omega^2}{2\sigma^2}\right] d\omega \quad (5)$$

where $x \equiv \omega - \omega_0$.

Defining $= \frac{1}{\sigma^2} - i3\tilde{b}\Omega$ , $B = \frac{\Omega}{\sigma^2} + i(2\tilde{a}\Omega - 3\tilde{b}\Omega^2)$, and $C = i(\tilde{b}\Omega^3 - \tilde{a}\Omega^2) - \frac{\Omega^2}{2\sigma^2}$, and using the Gaussian integral $R(\Omega) = \sqrt{\frac{\pi}{A(\Omega)}} exp\left(\frac{B(\Omega)^2}{4A(\Omega)} + C(\Omega)\right)$, we calculate the resulting Raman phase $\Phi_{Raman}(\Omega) = \arg(R(\Omega))$. This phase consists of two parts: the phase from the complex prefactor $\arg\left(\sqrt{\frac{\pi}{A}}\right) = \frac{1}{2}\arctan(3\tilde{b}\Omega\sigma^2)$ and the phase from the exponent $\frac{B(\Omega)^2}{4A(\Omega)} + C(\Omega)$. Added together (after arithmetics) we get:

$$\exp(i\phi_{Raman}(\Omega)) = \exp\left[i \cdot \left(\frac{\tilde{b}\Omega^3(9\tilde{b}^2\Omega^2\sigma^4 - 12\tilde{a}^2\sigma^4 + 1)}{4(9\tilde{b}^2\Omega^2\sigma^4 + 1)} + \frac{1}{2}arctan(3\tilde{b}\Omega\sigma^2)\right)\right] \quad (6)$$

To match the centrifugal phase term $\Phi_{CDN}$ one can extract the quadratic and cubic phase coefficients, $\tilde{a}$ and $\tilde{b}$, by brute-force minimization of the difference between $\Phi_{CDN}$ and $\Phi_{Raman}$ weighted by the Boltzmann population $\rho_J = (2J+1)e^{-E_J/kT}$ across all Raman frequencies ($\Omega$): $\min \sum_J \rho_J[\Phi_{Raman}(\Omega_J) - \Phi_{CDN}(\Omega_J)]^2$

An analytic approximation for the cubic phase $\tilde{b}$ is derived by equating the centrifugal phase term in (2) and the Raman phase of (6) excluding the arctan term:

$$\frac{\tilde{b}\Omega^3(9\tilde{b}^2\Omega^2\sigma^4 - 12\tilde{a}^2\sigma^4 + 1)}{4(9\tilde{b}^2\Omega^2\sigma^4 + 1)} = 4\pi n\frac{D}{B}(2J^3 + 9J^2 + 15J + 9)$$

Setting $\tilde{a} = 0$ and expressing the right-hand side in $\Omega \approx 2B(2J+3)$, we obtain the analytic approximation for the cubic phase coefficient:

$$\tilde{b} \cong \frac{\pi n D}{2B^4}$$

The above approximation for $\tilde{b}$ was found in very good agreement with the numerical brute-force least square method with <1% difference for $n > 5$. For practical experimental utilization, the difference between the brute-force optimization and the analytic approximation were found practically negligible. Thus, in what follows we present experimental results of CDN pre-compensation via cubic phase shaped pulses [1] using this analytic design.

Experimental results

**Validation in methyl-iodide gas**

Methyl iodide ($CH_3I$) is known for its significant centrifugal distortion with rotational constants $B = 0.25\ cm^{-1}$ and $D = 2.1 \cdot 10^{-7}\ cm^{-1}$ [29]. As such it has been the subject of coherent rotational control experiments [30] where the strong dephasing due to CDN served as a dephasing mechanism for rephasing echo schemes [14,16,31].

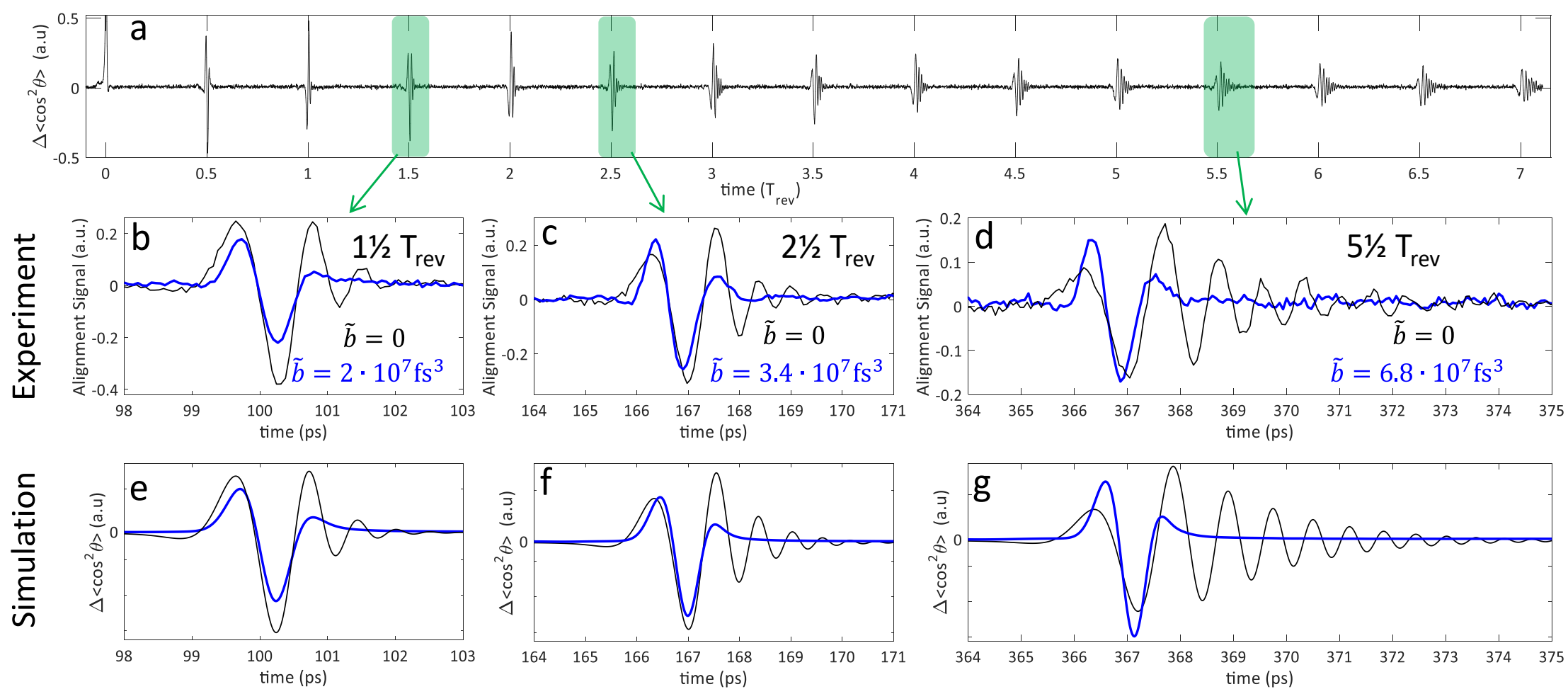

Figure 2 (a) Time-resolved birefringence signal from $CH_3I$ gas (15 torr, room temperature) excited by a 120fs, 800nm pulse (without phase shaping). Figs.(b,c,d) enlarged view of the 1.5, 2.5, 5.5 revival signals respectively. Black signals were induced by a transform limited pulse (unshaped, as in fig.2a) and blue signals were induced by cubic-phase-shaped pulses with $\tilde{b}$ values noted in the figures. Figs.(e,f,g) show the simulation results for (b,c,d) respectively.

Figure 2a depicts the time-resolved alignment signal of $CH_3I$ following excitation by a transform-limited (unshaped) near-IR pulse with $\sim 120$ fs duration. The effects of CDN on the shape of the revival signals manifests as gradual increase in the number of oscillations of the revival signals. Owing to the relatively large CDN of $CH_3I$ ($B/D \sim 1.2 \cdot 10^6$), these effects are observed already at $1.5T_{rev}$ ($\sim 100$ ps) as depicted by the black curve in Fig.2b. This distortion can be pre-compensated by the cubic-phase shaped pulse (with $\tilde{b} = 2 \cdot 10^7$ fs$^3$), depicted by the blue curve, with the typically single-cycle shape of the half-revival signal. At larger revival periods (longer delays), the CDN ramifications becomes more severe as shown by extended, multi-oscillatory signals shown by the black curves in Figs.2c,d. Respectively, pre-compensation for the larger distortion requires larger cubic phases with $\tilde{b} = 3.4 \cdot 10^7$ fs$^3$ and $\tilde{b} = 6.8 \cdot 10^7$ fs$^3$ for the $2.5\, T_{rev}$ (Fig.2c) and $5.5\, T_{rev}$ (Fig.2d) signals respectively. Figs.2e-g present the theoretical counterparts to Figs.2b-d, simulated using the same cubic phase parameters of the experiments and based on the analytical approximation derived above. The excellent agreement between experiment and theory validates the analytical treatment of the cubic phase as a reliable predictive design tool for CDN pre-compensation. We note that the time axes of the experimental and simulated data are determined by setting the first signal peak (end of the excitation pulses) at t=0.

**Direct observation of $l$-doubling alignment transients**

Leveraging our ability to mitigate centrifugal distortion effects at a chosen time, we now turn to resolve the rotational contributions of $l$-doubling to the alignment dynamics of $CO_2$. This approach allows for the direct time-domain observation of these rotational-vibrational contributions, previously masked by non-rigid rotor dynamics.

The rotational constants of $CO_2$ due to coupling with the perpendicular (bending) vibration are $B_{000} = 0.3902$, $B_{010}^{+} = 0.3912, B_{010}^{-} = 0.3905\, cm^{-1}$ [32] with respective revival periods of $T_{rev}^{000} = 42.74\, ps$, $T_{rev}^{010^+} = 42.63\, ps$, $T_{rev}^{010^-} = 42.71\, ps$. In order to resolve the alignment signals of the $l$-doubling components apart we set to probe them at a time where their selective alignment events are temporally separated by more than their signals' duration. The latter is $\sim 1ps$ as marked by the experimental alignment signal of $CO_2$ in Fig.3a. With the smallest difference among the revival periods of the three components being $\sim 30fs$, we set to resolve them around the $\left(\frac{1ps}{30fs}\right) \approx 33^{rd}$ revival period, at $t \approx 1400ps$ past excitation. At such long delays the alignment signals of $CO_2$ ($B \sim 0.39\ cm^{-1}, D = 1.33 \cdot 10^{-7}\ cm^{-1}$) [32] are severely affected by CDN with multiple oscillations and temporal elongation thwarting their distinction due to temporal overlap (Fig.3b). For better visibility of the $l$-doubling components associated with the v=1 bend ($\omega_{bend} = 667\ cm^{-1}$), we increased their Boltzmann population by heating the gas to $403K$ for which the populations are 84.6%, 7.7% and 7.7% for the $B_{000}, B_{010}^{+}$ and $B_{010}^{-}$ components respectively. Note that increasing the gas temperature acts as a double-sided sword; while the amplitude of the $l$-doubling components increase with temperature, the rotational distribution shifts to higher J levels resulting in more prominent distortion (since CDN$\propto J^4$) as observed in Fig.3b where the heated gas ( $403\ K$) is measured at $t \sim 1380$ ps following excitation by a 120 fs duration (unshaped) pulse.

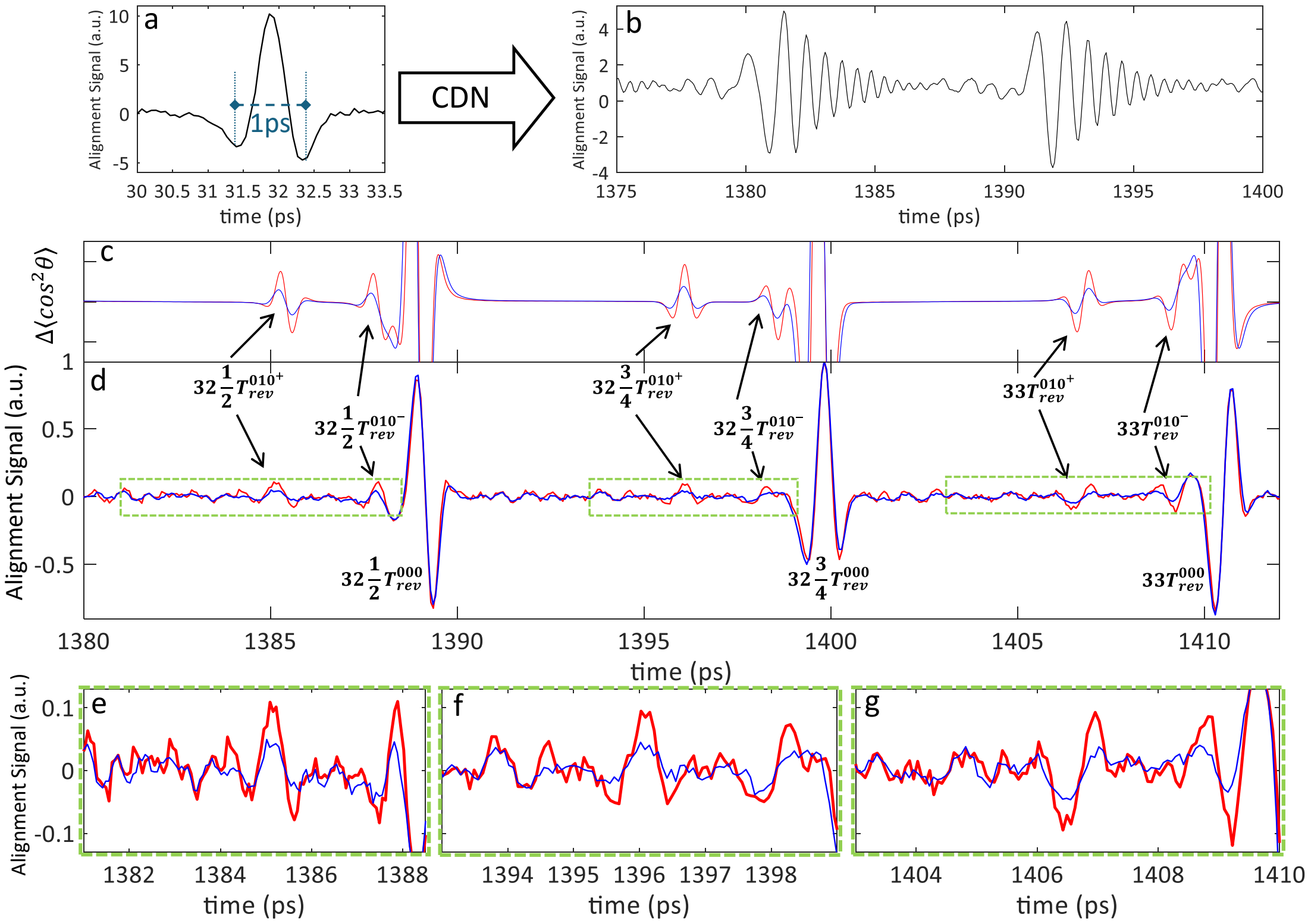

Figure 3 (a) Alignment signal induced by an unshaped 120fs excitation pulse in CO2 at 3/4 $T_{rev}$ and at (b) the 32$T_{rev}$ period showing severe multi-oscillatory signal shape. (c) Simulated CDN-free alignment signal at the 32$T_{rev}$ including the l-doubling contributions of CO2 at 293 K (blue) and 403 K (red). (d) Experimental revival signal at the 32$T_{rev}$ period of CO2 induced by a cubic phase shaped pulse at 293 K (blue) and 403 K (red). (e,f,g) enlarged view of selected regions marked by the dotted green rectangles in (d).

Fig.3d depicts the experimental alignment in the range 1380-1412 ps of $CO_2$ induced by a cubic-phase-shaped pulse, designed to pre-compensate for the CDN effects at the $32\mathrm{T}_{\mathrm{rev}}$ period. The blue and red curves were measured at 293 K and 403 K respectively and normalized to the $32\frac{3}{4}\mathrm{T}_{\mathrm{rev}}$ of the ground vibrational level ($\mathrm{B}_{000}$) selectively. The alignment signals associated with the three rotational components are marked in the figure, in agreement with the numerically simulated signals shown in Fig.2c. The latter were calculated by discarding the centrifugal distortion (D) in the energy term of the rotors, and provide further validation for the effective CDN-free result of Fig.2d.

We note the additional oscillations in the experimental signal background shared by the red and blue experimental curves in Figs.3d-g. These result from the phase shaping device (SLM) being pixelated resulting in satellite replicas in the shaped pulse and overall compromised pulse shape [33,34]. Naturally, as the required phase gradient increases, so does the distortion of the shaped pulse. Nevertheless, the molecular response proved remarkably robust against spectral phase discretization as can be appreciated from the experimental results.

**Conclusions**

In summary, we have experimentally demonstrated a robust scheme for the pre-compensation of CDN in rotational wavepacket dynamics of gas-phase molecules. By tailoring the spectral phase of an ultrafast excitation pulse with a cubic profile, we effectively negate the J-dependent dephasing that typically leads to the temporal elongation and multi-cycle fragmentation of rotational revivals. We further provide an analytical framework for pulse design, showing that the required cubic phase coefficient $\tilde{b}$ is directly related to the molecular constants with $\tilde{b} \approx \frac{\pi n D}{2B^4}$.

The capabilities of this technique are illustrated by the direct time-domain observation of $l$-doubling components in $CO_2$, a phenomenon that has been largely discarded in the realm of molecular alignment. Compression of CDN-distorted signals into near single-cycle revivals at long delays (t $\approx 1400$ ps) enabled resolving of distinct contributions from the ground state (v = 0) and excited bending modes (v = 1), which otherwise remain overlapped. Despite the finite resolution of pixelated pulse-shapers, the molecular response proved remarkably resilient, yielding high-fidelity signals that align with our theoretical models.

More broadly, this approach enables selective access to closely spaced rotational contributions as exemplified by $l$-doubling, and is directly applicable to molecular isotopologues. Such selectivity opens new possibilities for targeted interrogation and coherent control. Further improvements in phase purity may be achieved through propagation in higher-order dispersive media, supplemented by minor pulse-shaping corrections.

**Acknowledgments:** We thank Mr. Eran Rosen and the TAU chemistry machine shop team for producing the spectroscopic gas cell

**Funding:** The authors acknowledge the support of the Israel Science Foundation (1856/22) and the PAZI foundation.

**Notes:** The authors declare no competing financial interest.

**Data Availability Statement:** The data underlying this study is provided within the manuscript and Supporting Information. The raw data is available from the corresponding author upon reasonable request.